\documentclass[twocolumn]{aastex62}
\usepackage{natbib}
\newcommand\Msun{M_{\odot}}

\graphicspath{{./}{figures/}}

\submitjournal{ApJ}

\shorttitle{First Detection of an Over-Massive Black Hole Galaxy UHZ1}

\begin{document}

\title{First Detection of an Over-Massive Black Hole Galaxy UHZ1: Evidence for Heavy Black Hole Seed Formation from Direct Collapse}

\author[0000-0002-7809-0881]{Priyamvada Natarajan}
\affiliation{Department of Astronomy, Yale University, New Haven, CT 06511, USA}
\affiliation{Department of Physics, Yale University, New Haven, CT 06520, USA}
\affiliation{Black Hole Initiative, Harvard University, 20 Garden Street, Cambridge, MA 02138, USA}

\author[0000-0001-9879-7780]{Fabio Pacucci}
\affil{Center for Astrophysics $\vert$ Harvard \& Smithsonian, 60 Garden Street, Cambridge, MA 02138, USA}
\affiliation{Black Hole Initiative, Harvard University, 20 Garden Street, Cambridge, MA 02138, USA}

\author[0000-0001-5287-0452]{Angelo Ricarte}
\affil{Center for Astrophysics $\vert$ Harvard \& Smithsonian, 60 Garden Street, Cambridge, MA 02138, USA}
\affiliation{Black Hole Initiative, Harvard University, 20 Garden Street, Cambridge, MA 02138, USA}

\author[0000-0003-0573-7733]{\'Akos Bogd\'an}
\affil{Center for Astrophysics $\vert$ Harvard \& Smithsonian, 60 Garden Street, Cambridge, MA 02138, USA}

\author[0000-0003-4700-663X]{Andy D. Goulding}
\affiliation{Department of Astrophysical Sciences, Princeton University, Princeton, NJ 08544, USA}

\author[0000-0002-1697-186X]{Nico Cappelluti}
\affiliation{Department of Physics, University of Miami, Coral Gables, FL 33124, USA}

\begin{abstract}
The recent Chandra-JWST discovery of a quasar in the $z \approx 10.1$ galaxy UHZ1 reveals that accreting supermassive black holes (SMBHs) were already in place 470 million years after the Big Bang. The Chandra X-ray source detected in UHZ1 is a Compton-thick quasar with a bolometric luminosity of $L_{\rm bol} \sim 5\times 10^{45}\, {\rm erg\,{s^{-1}}}$, which corresponds to an estimated BH mass of $\sim 4\times 10^7\,\Msun$ assuming accretion at the Eddington rate. \textit{JWST} NIRCAM and NIRSpec data yield a stellar mass estimate for UHZ1 comparable to its BH mass. These characteristics are in excellent agreement with prior theoretical predictions for a unique class of transient, high-redshift objects, Over-massive Black Hole Galaxies [OBGs] by Natarajan et al., that harbor a heavy initial black hole seed that likely formed from the direct collapse of the gas. Given the excellent agreement between the observed multi-wavelength properties of UHZ1 with theoretical model template predictions, we suggest that UHZ1 is the first detected OBG candidate. Our assertion rests on multiple lines of concordant evidence between model predictions and the following observed properties of UHZ1: its X-ray detection and the estimated ratio of the X-ray flux to the IR flux that is consistent with theoretical expectations for a heavy initial BH seed; its high measured redshift of $z \approx 10.1$, as predicted for the transient OBG stage ($9 < z < 12$); the amplitude and shape of the detected JWST Spectral Energy Distribution (SED) between 1 - 5 microns, which is in very good agreement with simulated template SEDs for OBGs; and the extended  \textit{JWST} morphology of UHZ1 that is suggestive of a recent merger, also expected for the formation of transient OBGs. As the first OBG candidate, UHZ1 provides compelling evidence for the formation of heavy initial seeds from direct collapse in the early Universe.
\end{abstract}



\section{Introduction} 
\label{sec:intro}

The origin of the first supermassive black holes in the Universe remains an open question in astrophysics. Accounting for actively growing SMBHs with masses of 
$\sim10^9\,\Msun$ in detected $z\gtrsim6$ luminous quasars from light initial seeds has proven to be challenging given the time available for assembling their inferred high masses (see recent review by \cite{fan+2023} and references therein). Given that BHs are expected to grow via mergers and accretion over cosmic time \citep{HaehneltPNRees1998,Pacucci_2020_growth}, information about their initial seeding is expected to be erased. Therefore, directly accessing the highest redshift population offers the best prospects \citep{RicartePN2018, Pacucci_Loeb_2022} to constrain BH seeding models. 

A range of theoretical seeding prescriptions operating as early as at $z \sim 20-25$ classified broadly as “light” and “heavy” seeding models have been proposed as starting points to account for the formation of observed SMBHs \citep{Natarajan2011,Volonteri_2012,Woods_2019}. Light seeds are believed to be the remnants of the first generation of stars, the so-called Population III stars, that result in the production of initial BH seeds with $10-100\,M_{\odot}$ \citep{madau01}. 
 
 Heavy seed models, on the other hand, propose the formation of  $10^{4}\,-\,10^{5}\,M_{\odot}$ seeds in several possible ways. First, heavy seeds could result from the direct collapse of pre-galactic gas disks \citep{loeb94,volonteri05,lodato06,begelman06,LodatoPN2007}.

There are other promising proposed pathways and sites to form DCBH seeds at early epochs (i) from halos with supersonic baryon streaming motions relative to dark matter \cite{Stacy+2011,Schauer+2017}; (ii) in highly turbulent halos that are now thought to create the first quasars \cite{Latif+2022} and (iii) from major mergers of massive galaxies \cite{Mayer+2023,Mayer_Bonoli2019}. An additional pathway involves rapid, amplified early growth of originally light seeds that may end up in conducive cosmic environments, such as gas-rich, dense nuclear star clusters \citep{AlexanderPN2014}. 

The final stages of the formation of the DCBH have been explored in simulations upto the stage where supermassive stars (SMSs) are produced. Rapid baryon collapse at the center of atomically cooled halos has been demonstrated to produce SMSs \citep{Hosokawa+2013,Woods+2017,Haemmerle+2018,Herrington+2023} that are then predicted to collapse either by the post-Newtonian GR instability or from the depletion of core fuel at the end of post main sequence burning to form DCBHs. These simulations are yet to track the subsequent formation and growth of the BH seed and the stellar component in the host galaxy.

Rapid mergers of light remnants in the early nuclear star clusters could also lead to the formation of heavy seeds at high redshifts as proposed by \cite{Devecchi+2009} as well as the runaway collapse of nuclear star clusters as proposed by \citep{Davies+2011}. In addition to these more conventional theoretical seeding models, primordial black holes (PBHs, \citealt{PBHs_1971}) that form in the infant Universe have also been explored as potential candidates to account for the origin of initial seeds for SMBHs (see \citealt{Cappelluti+2022} and references therein), as well as dark stars that are postulated to be powered by dark matter self-annihilation in their cores \citep{Freese+2016}.

In this work, we focus on one scenario for DCBH formation in which viable DCBH formation sites in the Cold Dark Matter-dominated cosmogony are pristine atomic cooling halos, satellites bound to the first star-forming galaxies. In these satellite subhalos, gas cooling and fragmentation and hence star formation are suppressed as the more efficient molecular hydrogen coolant gets rapidly dissociated due to irradiation by the Lyman-Werner photons from the parent star-forming halo \citep{Wise+2008,Agarwal+2013,Regan+2017,Wise+2019,Patrick+2023}.

\begin{figure*}
\begin{center}   
\includegraphics[width=0.8\textwidth]{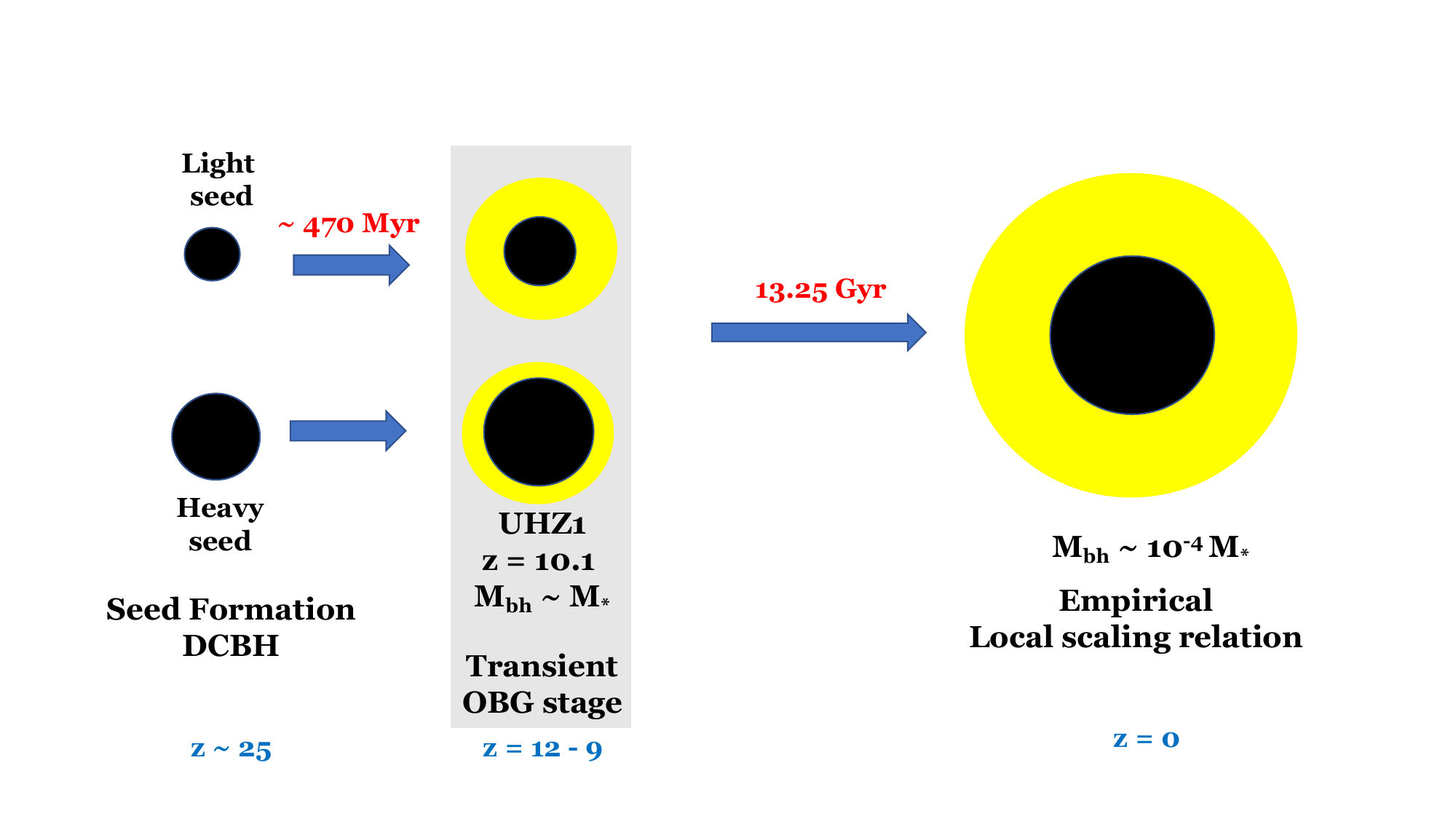}
\caption{Schematic diagram of the potential assembly history of the OBG candidate UHZ1. The direct collapse of primordial gas disks resulting in the production of heavy initial black hole seeds has been demonstrated to occur feasibly in the setting shown here: satellite DCBH halos that are bound to parent star-forming halos \citep[see Figure~1 in][]{Natarajan+2017}. Star formation in the satellite atomic cooling DCBH halo is expected to be suppressed due to the dissociation of molecular hydrogen from the Lyman-Werner radiation produced by the parent galaxy. As a consequence, the OBG is a merger remnant that contains a growing DCBH seed with an initial mass of $\sim 10^4\Msun$ with $M_{\rm bh} \geq {M_*}$}.
\label{fig:schematics}
\end{center}
\end{figure*}

The satellite DCBH subhalo is then predicted to rapidly merge within $\sim$ 1-5 Myrs with the parent star-forming halo to produce a new, transient class of high-redshift objects, Outsize Black hole Galaxies (OBGs) \citep{Agarwal+2013,Natarajan+2017}. The merger product, an OBG, would then harbor a growing central heavy BH seed procured from the satellite and the stellar population contributed by the parent galaxy. A schematic outline of the formation process of an OBG is shown in Figure~\ref{fig:schematics}. Post-merger, the stars and the BH would continue to grow self-consistently as the same gas reservoir that feeds the BH also forms stars. Given that $M_*\,\sim\, M_{\rm bh}$ for an OBG \citep{Natarajan2011,Agarwal+2013, Pacucci_2017,Visbal_2018}, this results in a strikingly different BH-to-host galaxy stellar mass ratio than observed in the local Universe, where the mass of the central BH is $\sim0.1\%$ of the stellar mass \citep{FerrareseMerritt2000,Tremaine+2002}. 

At present, this is the only scenario in which the entire arc of the formation, early growth of the DCBH seed and properties the stellar population of the host galaxy and its observational consequences in terms of high-redshift multi-wavelength signatures via spectral predictions have been predicted. For the other proposed channels for DCBH formation, the detailed exploration of the relationship between the mass of the growing DCBH seed and that of the stars in the host galaxy is awaited and the nature of this relation may well provide the discrimination between the multiple DCBH formation channels. For instance, DCBHs resulting from the major mergers of massive galaxies at extremely high redshifts as proposed by \cite{Mayer_Bonoli2019} are unlikely to lead to the formation of OBGs due to the abundance of stars available in these merger remnant hosts.

Data flowing currently from the \textit{JWST} is rapidly reshaping our understanding of early galaxy formation with the reported detection of large numbers of faint, distant galaxies at $z>9$ \citep{castellano22a,castellano22b,harikane22,naidu22,atek23,adams23,Leung+2023,Atek+2023}, many of which may harbor central BHs \citep{bogdan2023+,Larson+2023,Maiolino+2023,EPOCH+2023}. 
 
Utilizing the achromatic nature of gravitational lensing, \citet{bogdan2023+} innovatively deployed \textit{Chandra} also to study the Abell~2744 field in X-ray wavelengths to detect magnified faint background galaxies and their accreting central BHs. Abell~2744 is a Frontier Fields cluster lens at $z\sim 0.31$ with an extremely well-calibrated lensing mass model.\footnote{See, for instance, the six independently derived lensing mass models for this Frontier Fields cluster that are \href{https://archive.stsci.edu/prepds/frontier/lensmodels/}{publicly available}.} \citet{bogdan2023+} report the $(4.2-4.4)\sigma$ detection of an X-ray emitting source in the $z\simeq 10.1$ galaxy, UHZ1. The \textit{JWST} image of UHZ1 appears to be extended, and its redshift has since been spectroscopically confirmed with \textit{JWST} NIRSpec data as $z \simeq 10.1$ \citep{Goulding+2023}. \citet{bogdan2023+} report a best-fit column density of $N_{\rm H} \approx 8^{+\inf}_{-7} \times 10^{24}$~cm$^{-2}$ and a corresponding intrinsic $2-10$~keV luminosity of $L_{\rm X,int} \approx 9 \times 10^{45}  \ \rm{erg \ s^{-1}} $ after correcting for the $\mu=3.81$ lensing magnification factor at the location of UHZ1. Taken together, these suggest the presence of an obscured, likely Compton-thick, accreting BH in UHZ1. 

Given the properties of the unique simultaneous \textit{Chandra} and \textit{JWST} detection for UHZ1, we make the case that UHZ1 is the first detected OBG candidate. We demonstrate this by comparing the multi-wavelength UHZ1 data in hand with theoretical OBG model template predictions previously reported in \cite{Natarajan+2017}. We note that its X-ray detection is what uniquely sets UHZ1 apart in contrast to all the other recent \textit{JWST} detections of high redshift accreting BHs.

The outline of this paper is as follows: in Section~2, we briefly describe the key ingredients of the models and the procedure used to determine multi-wavelength template spectra for OBGs; we present the comparison of UHZ1 data with multi-wavelength model predictions of growing early DCBH seeds in Section~3. We conclude with a discussion of the implications of this first detection of an OBG for a deeper understanding of BH seeding channels and the assembly history of early black holes in Section~4.

\section{Derivation of OBG model templates}

We extended and expanded the tracking of early BH seed growth done using the GEMS code presented in \cite{Pacucci_2015_shining}, \cite{Pacucci_2016_firstdcbh,Pacucci+2017}, to include the cosmological context of the OBG stage as shown in the schematic in Figure~1 and described in \cite{Natarajan+2017}. We constructed a library of template multi-wavelength spectral models by varying the following key parameters while tracing the formation and evolution of the OBG: (i) the metallicity of the gas and stellar population; (ii) the accretion mode - the standard radiatively efficient thin disk mode (Eddington limited as it is feedback regulated) and the radiatively inefficient slim disk (super-Eddington accretion is permitted as the accretion is entirely gas supply limited in that instance) and (iii) age of the stellar population. The range of initial conditions for these cases, including gas fraction are adopted from the First Billion Years (FiBY) project cosmological simulation, wherein both DCBH formation and light seed formation sites are selected. Based on the SPH code Gadget, FiBY contains an equal number of gas and dark matter particles ($684^{3}$ each; in a box with a 4 comoving Mpc on a side; with gas particle mass of $1253 \Msun$ and dark matter particle mass of $6161\Msun$ which permits atomic cooling halos to be well resolved and studied). Further details of the FiBY simulation including prescriptions adopted for star formation, and Lyman-Werner radiation thereform and feedback are reported in \cite{Agarwal+2014}. Initial conditions adopted for feasible BH seed formation sites from fiBY are adopted for the GEMS runs.

The library of model templates for OBGs is generated by simultaneously following the evolution of growing BH seeds (heavy and light) and the accompanying stellar populations with a range of ages and metallicities. The early growth of the BH seed is computed separately using 1-dimensional radiation-hydrodynamical models \citep{PacucciFerrara2015, Pacucci_2015_shining, Pacucci_2016_firstdcbh,Pacucci+2017}. These models simulate spherical accretion onto the high-redshift seed BH, and calculate the emitted luminosity self-consistently from the mass accretion rate. The key input parameters for these models are the gas density and metallicity (taken from the fiBY cosmological simulation) and the BH seed mass. The post-processing spectral analysis is done using CLOUDY \citep{Ferland+2013}. 

\begin{figure}
\begin{center}
    \includegraphics[width=0.48\textwidth]{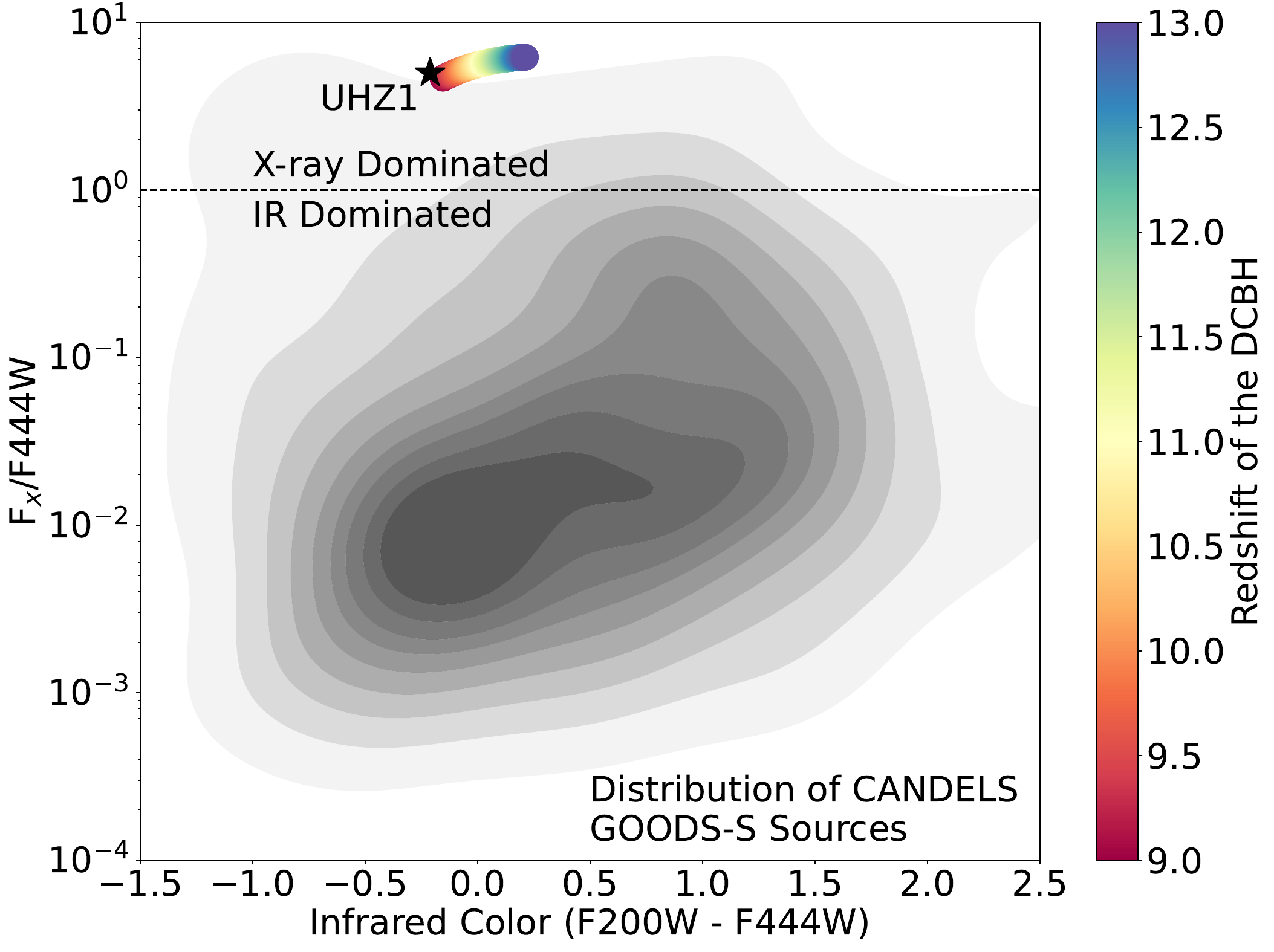}
\caption{{Selection criteria for OBGs: the predicted location of OBGs in color-color space, with the location of UHZ1 marked with a black star.}}
\label{fig:colors}
\end{center}
\end{figure}

For our models the gas fraction in viable DCBH formation sites and the star formation history of the parent halo are are taken from fiBY \citep{Agarwal+2014}. The evolution of the accompanying stellar component is tracked simultaneously and finally combined with that of the growing BH. 

Two limiting cases for growth by accretion are implemented to produce a library of models \citep{Pacucci_2015_growth_efficiency}: standard accretion - which adopts the standard $\alpha$-disk model that is geometrically thin and optically thick, and hence radiatively efficient with accretion capped at the Eddington rate and slim disk accretion, which is characterized by a geometrically thick disk, that is radiatively inefficient where radiation pressure is less efficient quenching gas inflow due to radiation trapping permitting super-Eddington accretion rates. The spectral shape of the output luminosity from accretion onto the BH is determined largely by the geometric properties of the accretion disk. Eddington limited accretion is the hallmark of thin-disk accretion during which the output luminosity $L_{\rm acc} \propto \dot M_{\rm acc}$, where $\dot M_{\rm acc}$ is the mass accretion rate. In this highly radiatively efficient regime, the luminosity is feedback limited. Meanwhile, as slim accretion disks result in super-Eddington accretion rates, in this instance, the output $L_{\rm acc} \propto [\ln {\dot M_{\rm acc}}]$. In this radiatively inefficient regime, gas accretion is expected to be supply limited. These two distinct geometries result in dramatically different fluxes mapped out in our models. Post-processing in CLOUDY utilizes inputs from GEMS at various time-slices for light and heavy seeds and includes obscuration. As the accreting BH and stellar component are evolved separately, feedback coupling between them is not taken into account in these models.

The combined contribution of fluxes from the stellar component and both light and heavy accreting seeds is computed in the generated synthetic models derived by combining the pure seed mass evolution followed via GEMS along with the stellar population. The stellar population is modeled using two possibilities: a younger, lower metallicity $(5\times 10^{-4}\,Z_{\Msun})$ and an older, higher metallicity population $(5 \times 10^{-2}\,Z_{\Msun})$, both modeled with a Kroupa IMF. Distinct seeding signatures are seen in the emergent spectrum. The resulting properties for the full parameter space comprising the two seeding scenarios, two accretion models, and two distinct assumptions for the metallicity of the stellar population are presented in detail in \citet{Natarajan+2017}. 

Robust selection criteria for OBGs powered by initially heavy seeds have also been derived, including a pre-selection to eliminate blue sources, followed by color-color cuts $([{F}090W-{F}220W]> 0$;$-0.3< [{F}200W-{F}444W]< 0.3$) and the ratio of X-ray flux to rest-frame optical flux $({F}X/{F}444W\gg 1)$. These cuts sift out OBGs from other bright, high- and low-redshift contaminants in the infrared. OBGs were predicted to have faint but detectable magnitudes of ${M_{\rm AB} < 25}$ and unambiguously detectable by the NIRCam (Near-Infrared Camera) on \textit{JWST}. Fainter growing light seed remnants with lower birth masses were found to have significantly fainter predicted AB magnitudes of ${M_{\rm AB}< 31}$ by $z \sim 10$.

\begin{figure}
\begin{center}
\includegraphics[width=0.48\textwidth]{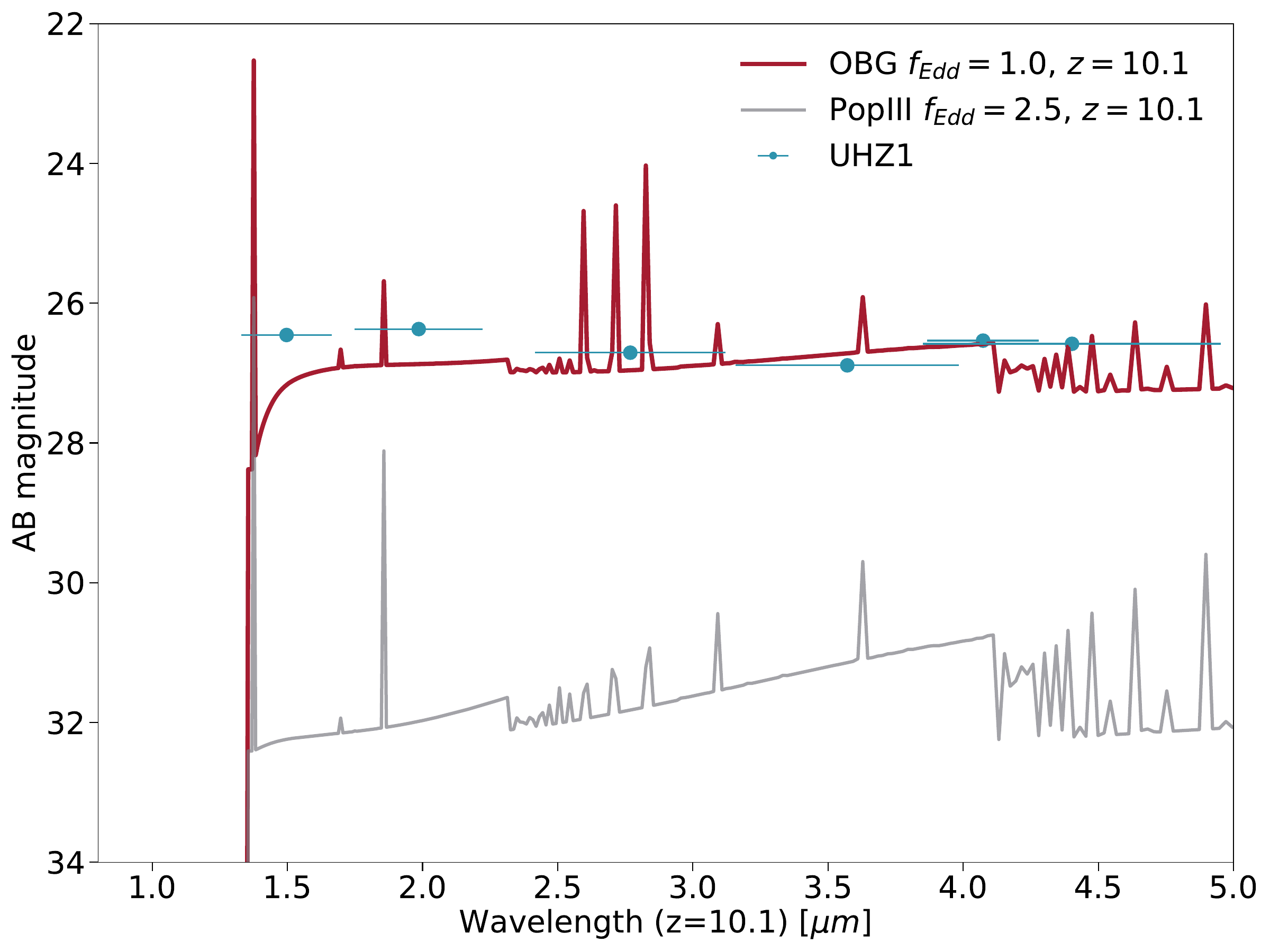}
\caption{The OBG model match for UHZ1: the model spectrum over-plotted here is obtained by growing an initially heavy seed of $\sim 10^4\,\Msun$ to a final mass of  $10^7\,\Msun$ as estimated for UHZ1 and combining with a young stellar population (age of 350 Myr), low metallicity ($10^{-3}\,Z_{\Msun}$) with a column density of $\sim 3 \times 10^{24}\,cm^{-2}$. The observed near-IR JWST SED for UHZ1 points taken from \cite{castellano22b} are shown in blue.  We note that the overplotted SED template from our library generated with the parameters noted above is very similar to UHZ1.}
\label{fig:sed}
\end{center}
\end{figure}

\section{Comparison of UHZ1 with OBG model templates}

We use the following observed properties of UHZ1 to find a template match from our library of early seed growth models first presented in \cite{Natarajan+2017}. As reported by \cite{castellano22a,castellano22b}, UHZ1, magnified by the foreground cluster Abell 2744, has an intrinsic magnitude that renders it detectable with a magnitude of $M_{\rm AB} \sim 27$ and a nearly flat SED in the observed \textit{JWST} bands spanning $1 - 5$ microns. Fitting a Salpeter IMF, \citet{castellano22b} infer a stellar mass for UHZ1 of $\sim 4\times 10^7\,\Msun$, and it is found that combining photometry with \textit{JWST} NIRSpec data \cite{Goulding+2023} provides a slightly higher stellar mass estimate. An independent fit performed by \cite{Atek+2023} also report a stellar mass of $\sim 7 \times 10^7\,\Msun$, making all these estimates broadly consistent with each other. The measured bolometric luminosity from \textit{Chandra} is $L_{\rm X}\sim 5\times10^{45}\ \rm{erg\ s^{-1}}$, yielding a BH mass of $\sim 4\times10^{7 -8} \ \rm{M_{\odot}}$ assuming accretion at the Eddington rate \citep{bogdan2023+}. In addition to these multi-wavelength data, the composite \textit{JWST} image of UHZ1 appears extended. We note that in their SED fit \citet{castellano22b} were unaware of the presence of an accreting BH. The X-ray detection of UHZ1, combined with \textit{JWST} flux measurements and recent \textit{JWST} NIRSpec data, motivates and strongly justifies our exploration of an OBG model match to UHZ1.

We also note the following uncertainties in interpreting the data for UHZ1. As the column density is weakly constrained, the mass estimate for the BH in UHZ1 from X-ray data is also not tightly constrained. Meanwhile, the stellar mass estimate from SED fitting to \textit{JWST} data also represents a lower limit as current templates do not fully take into account the underlying older stellar population that may be contained in sources like UHZ1. 

We note that the current \textit{JWST} SED fitting for UHZ1 adopted by \citet{castellano22b} assumes that the observed UV/optical emission derives solely from the stellar component modeled with a Salpeter IMF. This fit was done before the detection of the accreting BH, which is reported to be heavily obscured by \citet{bogdan2023+}. With the subsequent knowledge and derived properties of the SMBH hosted in UHZ1 from the X-ray data, we explore whether the rest-frame UV/optical SED and corresponding X-ray flux detected are compatible with our theoretical model templates of OBGs \citep{Natarajan+2017}.

We first explore the mass accretion history of the BH in UHZ1, starting from initially light ($100\,\Msun$) and heavy seeds ($10^4\,\Msun$) at $z > 20$ to reach the final inferred BH mass of $\sim 10^{7}\,M_{\odot}$. An initially light seed with a birth mass of $10-100\,M_{\odot}$, would need to steadily accrete at well above the Eddington rate (over 2 $\times$ the Eddington rate) for roughly 300 million years, while a heavy DCBH seed with an initial birth mass of $10^{4}\,M_{\odot}$, could reach the final mass of the BH powering UHZ1 while accreting at just the Eddington rate throughout.  

Next, we look at template model matches from our library for the multi-wavelength SEDs for UHZ1 for the two possible seeding scenarios; the two assumed accretion models and possibilities for the age and metallicity of the stellar population that bracket the entire permitted parameter space. Our template library includes outputs for the time slice with the largest detectable X-ray flux.

To explain UHZ1 with a light initial seed of $100\,\Msun$, accreting at super-Eddington rates captured via our slim disk accretion model, we do not find a template match that simultaneously satisfies \textit{JWST} and \textit{Chandra} data for UHZ1. A template that comes somewhat close is the predicted model \textit{JWST} SED is shown in Figure~\ref{fig:sed} (grey spectrum) where it is seen that the predicted \textit{JWST} flux is approximately two orders of magnitude lower than observed, resulting in the production of a significantly fainter source, predicted to have $M_{\rm AB} \leq 31$, at the sensitivity limit of \textit{JWST}. Additionally, UHZ1 would have been undetected in X-rays in contradiction to what is seen. Therefore, the simultaneous X-ray and \textit{JWST} detection of UHZ1 given the final BH mass of $\sim 10^7\,\Msun$ that is needed to be in place by $z=10.1$ strongly disfavors UHZ1 as originating from a light seed. 

Meanwhile, for a heavy seed origin model for UHZ1 with a seed mass of $10^{4}\,M_{\odot}$, we do find a template match from our library that simultaneously satisfies \textit{JWST} and \textit{Chandra} data for both assumed high and low metallicity values for the stellar population for a range of permitted stellar ages, as there is a trade-off between these two attributes. 

\begin{figure}
\begin{center}   
\includegraphics[width=0.48\textwidth]{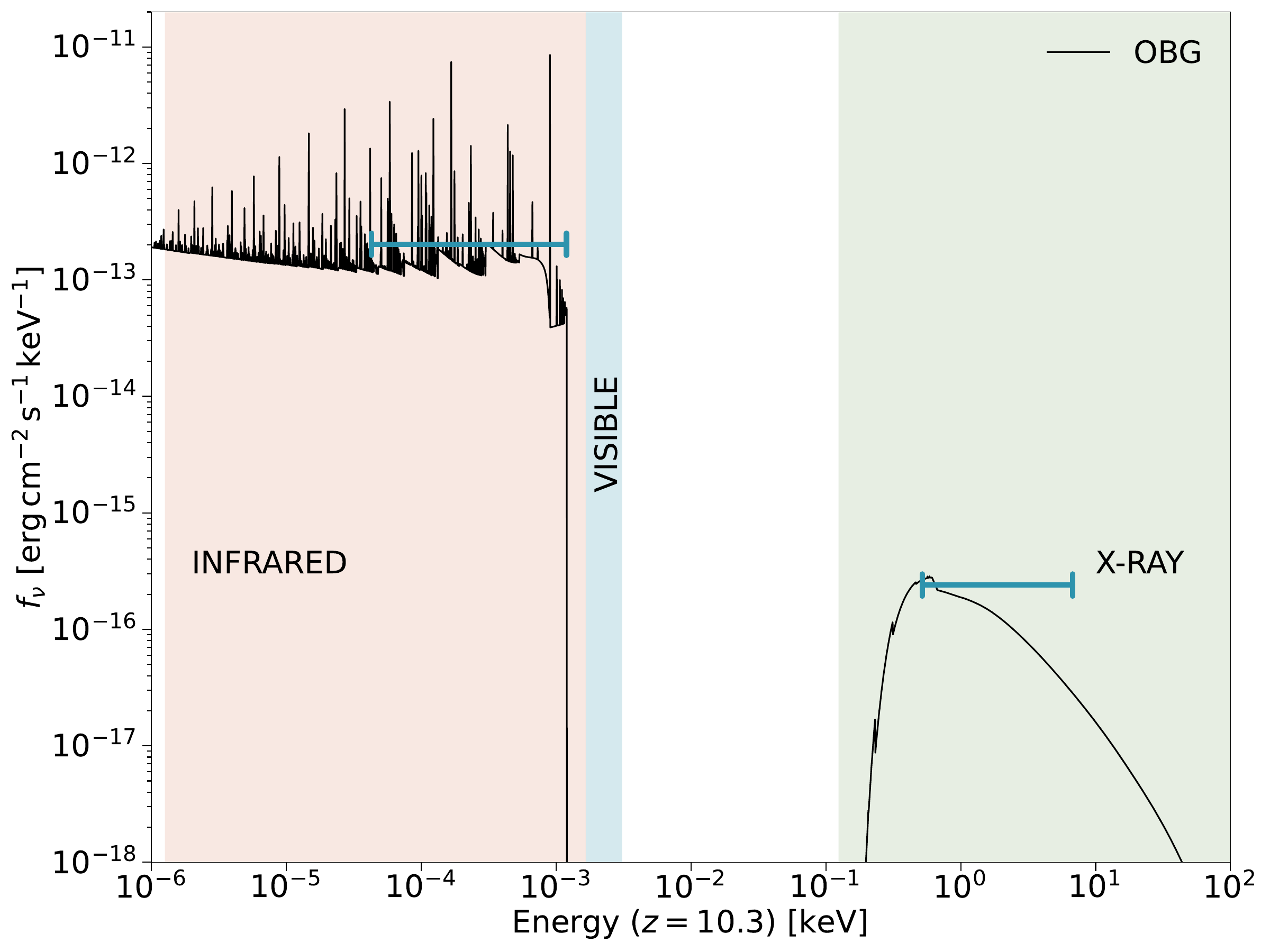}
\caption{{The template match for the multi-wavelength SED from our library of OBG models on which we over plot the measured fluxes in the IR from \textit{JWST} and X-ray from \textit{Chandra} for UHZ1, shows that they are consistent and well matched. The template that provides this optimal match has the following properties: a heavy initial seed with a mass of $10^{4}\,M_{\odot}$; the OBG accreting at the Eddington limit with an age of 350 Myr for the recently merged stellar population with a low metallicity of $\sim\,10^{-3}\,Z_{\odot}$} and a column density of $3 \times 10^{24}\,cm^{-2}$, shown in Figure~\ref{fig:sed} (in red).}
\end{center}
\label{fig:multi_sed}
\end{figure}

For models where growth is via standard Eddington-limited accretion with a column density of $N_H \sim 3 \times 10^{24}\,cm^{-2}$: (i) the predicted hard X-ray flux is consistent with the measured value compatible with the inferred column density as shown in Figure~\ref{fig:multi_sed}; (ii) the flux ratio ${F_{444}/F_X} \sim 1$ as expected for OBGs, shown in Figure~\ref{fig:colors} and (iii) UHZ1 satisfies all the color-color selection criteria for an OBG, also seen in Figure~\ref{fig:colors}. The predicted \textit{JWST} SED shape and amplitude from this template with an age of 350 Myr for the recently merged stellar population with a low metallicity of $\sim\,10^{-3}\,Z_{\Msun}$ matches the data very well as shown in Figure~\ref{fig:sed} (red spectrum).  

In contrast, we note that for templates with a heavy seed of mass $10^{4}\,M_{\odot}$ that subsequently grows via slim disk accretion at potentially super-Eddington rates limited by the available gas supply, the OBG stage is simply too short-lived. While it could be potentially detectable with \textit{JWST} for the higher metallicity case (for the stellar population in the host galaxy) due to the lowered X-ray flux from the extremely strong obscuration, such sources would be undetected even with the deepest current X-ray exposures. Besides, with an extremely short predicted lifetime in the OBG stage of $\sim$ 5-10 Myrs, these sources would rapidly transit toward $M_* > M_{\rm bh}$. Once again, our \textit{Chandra} X-ray detection of UHZ1 rules out this family of templates. We map out the parameter space of template models and tabulate the possibilities for UHZ1 in Table~1.

\begin{table*}[ht]
\centering
\begin{tabular}{c l l r r}
\hline
{\bf Metallicity}&{}&{\bf Accretion Mode}&{}&{}\\
{\bf of}&{}&{}&{}&{}\\
 {\bf Stellar Component}&{Thin Disk}&{Thin Disk}&{Slim Disk}&{Slim Disk}\\
 \hline
 {} & {\it Chandra} & {\it JWST} & {\it Chandra} & {\it JWST} \\
 \hline
 {Low Z} & {\checkmark} & {\checkmark} & {X} & {\checkmark} \\ 
 {High Z} & {\checkmark} & {\checkmark} & {X} & {\checkmark}
 \\
 \hline
 \end{tabular}
\caption{Synopsis of multi-wavelength detectability for an initially {\bf heavy seed} of $10^4\Msun$ from our model template library of predicted SEDs given the current sensitivity limits of \textit{JWST} and the deepest available current \textit{Chandra} data for a source with the observed properties of UHZ1 at $z \sim 10$.}   
\end{table*}

\begin{table*}[ht]
\centering
\begin{tabular}{c l l r r}
\hline
{\bf Metallicity}&{}&{\bf Accretion Mode}&{}&{}\\
{\bf of}&{}&{}&{}&{}\\
 {\bf Stellar Component}&{Thin Disk}&{Thin Disk}&{Slim Disk}&{Slim Disk}\\
 \hline
 {} & {\it Chandra} & {\it JWST} & {\it Chandra} & {\it JWST} \\
 \hline
 {Low Z} & {X} & {at the limit} & {X} & {X} \\ 
 {High Z} & {X} & {at the limit} & {X} & {X}
 \\
 \hline
 \end{tabular}
\caption{Synopsis of multi-wavelength detectability with for an initially {\bf light seed} from our model template library of predicted SEDs given the current sensitivity limits of \textit{JWST} and the deepest available current \textit{Chandra} data for a source with the observed properties of UHZ1 at $z \sim 10$.}   
\end{table*}

\section{Conclusions \& Implications for early BH seeding}

Despite the theoretical uncertainties in our current understanding of early BH seeding, the detection of even a single high-redshift X-ray quasar, UHZ1, at $z \approx 10.1$ has significant implications as it provides new empirical information on the properties of initial BH seeds and coupling of accreting early BHs and their host galaxies. 

With this unique multi-wavelength observational data-set for UHZ1, we present compelling evidence that it represents the first detection of an OBG, the class of high-redshift galaxies that are seeded with heavy initial BHs, as predicted by \citet{Natarajan+2017}. As we have shown, all OBG selection criteria are satisfied by UHZ1 with an initial heavy seed mass of $10^4\,\Msun$, and its growth history is compatible with standard Eddington-limited accretion. The best model match from our OBG template library for UHZ1 is therefore provided by a heavy initial seed with a low metallicity for the stellar component and age of $\sim$ 300 Myr. There is a trade-off between the age and metallicity of the stellar population in OBGs. This degeneracy is well-documented more generally for stellar populations at other cosmic epochs as well. We, therefore, claim that UHZ1 offers compelling empirical evidence for the existence of heavy seeds in the early Universe. 

Studies of the BH seeding epoch have thus far been restricted to theoretical explorations. Forming BH seeds ab-initio in cosmological simulations and tracking their growth history is extremely challenging. Most simulation suites adopt simple seeding prescriptions to track BH growth and feedback over cosmic time. Detecting the first OBG candidate, UHZ1, offers empirical guidance to inform our heavy seeding theoretical models. As JWST progressively brings the $z>10$ Universe into view, detecting individual extreme accreting BHs at the earliest epochs is likely to provide new powerful insights in the coming years. 

In this work, we are explicitly not making a case for extrapolation of the growth of UHZ1 down to $z = 6-7$. We do not claim it is a likely progenitor for the luminous optically detected SDSS quasars. We are cautious about this as cosmological simulations, the {\sc MASSIVEBLACK} suite in particular, have shown that the most massive BH at $z \sim 10$ does not necessarily remain and grow to be the most massive BH by $z = 6$ \citep{DiMatteo+2017, DiMatteo+2023}. We emphasize that the details of the larger scale environment play an important role in shaping the accretion and, therefore, the growth history of BHs. Neither our 1-D hydrodynamical BH growth tracking simulations, whose results informed our model templates, nor large cosmological boxes at present adequately capture gas flows in the larger environment of BHs. Conversely, in our current analysis, we are agnostic to the Eddington ratio distributions inferred at lower redshifts for observed X-ray AGN.

We also refrain from making any number density estimates for OBGs relying on the detection of a single source as models and simulations \cite{Wise+2019,Regan+2020,Whalen+2020} indicate that DCBHs are rare and may be expected to account only for a small fraction of the most luminous quasars detected at $z \sim 6-7$. Estimates of the predicted relative abundance of light and heavy seeds are currently highly uncertain. Lack of information on the occupation fraction of BH seeds and an incomplete census of X-rays from early accretors due to obscured accretion preclude detailed calculations at present. However, a rough back-of-the-envelope estimate given the currently observed \textit{JWST} fields (UNCOVER, GLASS, \& CEERS surveys) and the detection of UHZ1 as the single OBG candidate suggests agreement with the expected number densities at $z \sim 10$ from just the number density of DCBH sites estimated in \citep{Natarajan+2017} from simulations of $\sim 10^{-6} - 10^{-7}$ per Mpc$^{-3}$. 

For accreting high-redshift BHs detected only by \textit{JWST} so far that are not necessarily OBGs and remain undetected in the deepest X-ray data in hand, one theoretical number density estimate has been attempted using semi-analytic models that we caution are also unable to capture the properties of the environment. \cite{Trinca+2023} discuss the expected number density of $z > 10$ AGN in \textit{JWST} fields with footprints and depths similar to that of UNCOVER/GLASS surveys that probe the Abell 2744 field and arrive at the following estimate for expectations: for CEERS-like survey $\sim$ 8 to 21 AGN at $7 \leq z \leq 10$; JADES-Deep about 12 to 63 AGNs with $7 \leq z \leq 10$ and 5 to 32 AGN at $z \geq 10$. As we note in this work, only a small fraction of these high redshift AGN will be OBGs and be detected in X-rays as well.

Our simple growth models certainly have limits that translate directly into the range of possibilities that we have mapped out to create our model template library as shown in Table~1. We have made simplifying assumptions - as this is the best we can do now as no current cosmological simulations can form BH seeds ab-initio and track their growth and that of their galaxy hosts self-consistently, taking the overall environment into account. 

As \textit{JWST} detects more $z > 9$ accreting BHs in the coming cycles, we plan to analyze those sources, investigate possible X-ray counterparts with \textit{Chandra}, and develop a deeper understanding of OBGs and heavy seeding physics. 

\section*{Acknowledgments}

PN acknowledges support from the Gordon and Betty Moore Foundation and the John Templeton Foundation that fund the Black Hole Initiative (BHI) at Harvard University where she serves as one of the PIs. F.P. acknowledges support from a Clay Fellowship administered by the Smithsonian Astrophysical Observatory. F.P. and A.R also acknowledge support from the BHI. A.B. acknowledges support from the Smithsonian Institution and the Chandra Project through NASA contract NAS8-03060 A.D.G. acknowledges support from NSF/AAG grant 1007094.

\bibliography{obj_letter}



\end{document}